\pdfoutput=1 
\documentclass{JINST}
\usepackage{upgreek}
\usepackage{cite}

\title{Beam energy determination in experiments at electron-positron colliders}

\author{M.N. Achasov$^{a,b,}$\thanks{Corresponding author.} ~ and
 N.Yu. Muchnoi$^{a,b}$\\
\llap{$^a$}Budker Institute of Nuclear Physics, Siberian Branch of the Russian
 Academy of Science,\\
 11 Lavrentyev, Novosibirsk 630090, Russia\\
\llap{$^b$}Novosibirsk State University,\\
 Novosibirsk 630090, Russia\\
E-mail: \email{achasov@inp.nsk.su}}

\abstract{The review of using of compton backscattering method for
determination of the beam energy in collider experiments is given.}


\begin{document}

\section{Introduction}\label{sec:xxx}
 
 The high accuracy of $e^+e^-$ colliders beam energy determination is crucial
 for a lot of physical studies:
\begin{itemize}
\item
particles masses and widths measurements,
\item
study of interference effects in the cross sections,
\item
measurements of the cross sections themselves.
\end{itemize}
For example, the error of the measured $Z$-boson mass $m_Z=91187.6\pm 2.1$ MeV
is dominated by the common LEP energy error 1.7 MeV \cite{pdg}. At the low
energy region in order to measure the $e^+e^-\to\pi^+\pi^-$ cross section
below 1 GeV with accuracy about 0.5\%, the beam energy should be measured
with relative error $\delta E/E\sim 10^{-4}$.

In the simplest case the beam energy can be calculated as
\begin{equation}
\label{eq:xxx}
E={B\rho\over\beta c}+\Delta_{cor},
\end{equation}
where $B$ and $\rho$ are magnetic field and beam radius of curvature,
$\Delta_{cor}$ is a nonlinear correction. In this case the relative accuracy 
of  the beam  energy determination $\delta E/E$ is worse than $10^{-3}$. The 
beam energy can also be determined by measuring of the momentum of particles 
in collinear events. For example, in special case of the process 
$e^+e^-\to\phi(1020)\to K^+K^-$, the beam energy
can been obtained with relative accuracy $5\times 10^{-5}$ at the 
$\phi$-meson peak \cite{kk} as
\begin{equation}
\label{eq:x1}
E=\sqrt{p_K^2+m_K^2}+\Delta_{cor},
\end{equation}
where $p_K$ is an average momentum of $K^+$ and $K^-$ mesons, $m_K$ is the 
charged kaon mass, $\Delta_{cor}$ is a correction due to kaon ionization 
energy losses and due to emission of real photons. The beam energy can be 
calibrated using positions of the narrow
and precisely measured resonances ($\omega,\phi,\psi,\Upsilon$).

Resonance depolarization (RD) method  is 
the most precise and has relative error of about $10^{-6}$ \cite{rd}.
The beam energy determination using RD requires the polarized beam and 
is based on the coupling between the electron energy and frequency 
$\Omega$ of its spin precession during the motion of the particle in
the transverse magnetic field with a revolution frequency
$\omega_s$:
\begin{equation}
E=\biggl(\frac{\Omega}{\omega_s}-1\biggr)\frac{\mu_0}{\mu^\prime}m_ec^2,
\label{depo}
\end{equation}
where $\mu_0/\mu^\prime$ is the ratio of the anomalous and normal parts of the
electron magnetic momentum known with a relative accuracy of $2\times
10^{-10}$ \cite{pdg}. The frequency $\Omega$ can be obtained through resonant
depolarization of the polarized beam due to impact of an external
electromagnetic field with a frequency  $\omega_d$ such that
\begin{equation}
\omega_d\pm k\omega_s = \Omega (k\in\mathbb{Z}).
\label{freq}
\end{equation}
Another possibility is the beam energy measurement using Compton
back-scattering (CBS) of monochromatic laser radiation on the electron beam.

\section{The Compton back-scattering approach}\label{sec:yyy}

CBS of laser light on electron beams is a well known method of generation of 
quasimonochromatic energetic photon beams. Let us consider the Compton 
scattering
process in a case when the angle $\alpha$ between initial particles is equal 
to $\pi$ and their energies are $\omega_0\ll m_e\ll E$. Here $\omega_0$ and
$E$ are the energies of the initial photon and electron, respectively. The
back-scattered photons have the maximal energy, and the energy spectrum of the
scattered photons has a sharp edge at the maximal energy:
\begin{equation}
\omega_{max}={{E^2}\over{E+m_e^2/4\omega_0}}.
\end{equation}
If one measures $\omega_{max}$, then the electron energy can be calculated:
\begin{equation}
E={\omega_{max}\over 2} \Biggl[ 1 +
\sqrt{1+{ m_e^2 \over {\omega_0\omega_{max}} } }\;\Biggr].\label{nepec}
\end{equation}

\begin{figure}[tbp]
\centering
\includegraphics[width=1.0\textwidth,height=0.6\textwidth]{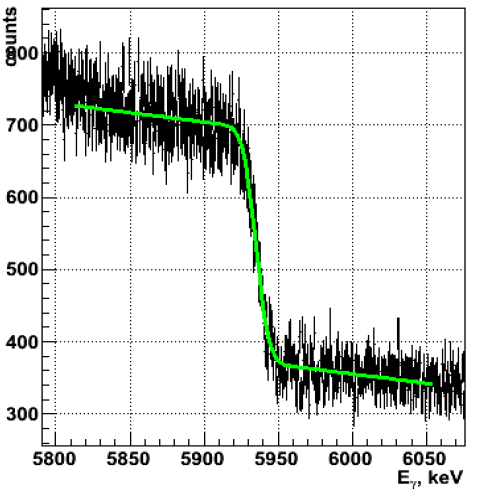}
\caption{The measured edge of the scattered photons energy spectrum at 
BEPC-II. Line is the fit result.}
\label{edge}
\end{figure}

The measurement procedure is as follows. The laser light is put in collision 
with the electron or positron beams, and the energy of the back-scattered
photons is precisely measured using the High Purity Germanium (HPGe) detector.
The maximal energy of the scattered photons is determined by fitting the abrupt
edge in the energy spectrum by the erfc-like function (figure~\ref{edge}). 
The beam energy $E$ is calculated from the maximum energy $\omega_{max}$ using
eq.~(\ref{nepec}).
CBS method provides rather high accuracy, a possibility of measurements in a 
wide energy region, allows to measure beam energy during data taking.

This approach was developed and experimentally proved in 
ref.~\cite{okp1,okp2,tls,bessy1-1,bessy1-2,bessy2,vepp4,bems,ems2000}.
First measurement of the beam energy $E\approx 1.3$ GeV at storage ring of 
Taiwan Light Source with accuracy $\delta E/E\sim 10^{-3}$ was reported in 
ref.~\cite{tls}. The accuracy of the measurements was improved at SR sources
BESSY-I and BESSY-II.  The relative accuracies of energy measurement of about 
$2\times 10^{-4}$ and $3\times 10^{-5}$  was achieved for the beam energy of 
800 MeV and  1.7 GeV at the BESSY-I and BESSY-II storage rings respectively
\cite{bessy1-1,bessy1-2}. The accuracies was proved by comparison with results
of RD method. In collider experiments, the CBS method was applied at 
VEPP-4M \cite{vepp4} and the $\tau-charm$ factory BEPC-II \cite{bems} for 
measurement of beam energy 1 -- 2 GeV, and at VEPP-2000 below 1 GeV
\cite{ems2000}.

\section{Beam energy measurement system for CBS method.}

A beam energy measurement system (BEMS) includes:
\begin{itemize}
\item
laser and optical system to provide initial photons and their transportation,
\item 
laser-to-vacuum insertion system which provides insertion of the laser beam 
into the vacuum chamber of collider,
\item
HPGe detector to measure backscattered photons spectrum,
\item
data acquisition system,
\item
data processing.
\end{itemize}

\subsection{HPGe detector}

The purpose of a HPGe detector is to convert gamma rays into electrical
impulses which can be used with suitable signal  processing, to determine
their energy and intensity. A HPGe detector is a large germanium diode 
operated in the reverse bias mode. At a suitable operating temperature 
(normally $\simeq$100 K), the barrier created at the junction reduces the 
leakage current to acceptably low values. Thus an electric field can be 
applied that is sufficient to collect the charge carriers liberated by
the ionizing radiation.

Commercially available coaxial HPGe detector with diameter 5 -- 6 cm and
height 5 -- 7 cm has energy resolution $\delta\omega/\omega\sim 10^{-3}$ and 
can measure the $\gamma-$quanta energy below 10 MeV. The detector is
connected to the multi-channel analyzer (MCA).

The statistical accuracy of beam energy measurement of about
$5\times (10^{-4} - 10^{-5})$ can be achieved in a reasonable time (about 1 
hour). The systematic accuracy is mostly determined by absolute
calibration of the detector, which could be
done in the photon energy range up to $\sim  10$ MeV by using the 
$\gamma-$active radionuclides (table~\ref{tab1}).

\begin{table}[tbp]
\caption{Radiative sources of $\gamma$-quanta which can be used for the HPGe
detector calibration.}
\label{tab1}
\smallskip
\centering
\begin{tabular}{|lc|}
\hline
Source & $E_\gamma$, keV \\
$^{208}Tl$  &$583.191 \pm 0.002$ \\
$^{137}Cs$  &$661.657 \pm 0.003$ \\
$^{60}Co$   &$1173.237 \pm 0.004$ \\
$^{60}Co$   &$1332.501 \pm 0.005$ \\
$^{208}Tl$  &$2614.553 \pm 0.013$ \\
$^{16}O^*$  &$6129.266 \pm 0.054$\\
\hline
\end{tabular}
\end{table}

\subsection{Laser}

\begin{figure}[tbp]
\centering
\includegraphics[width=1.0\textwidth]{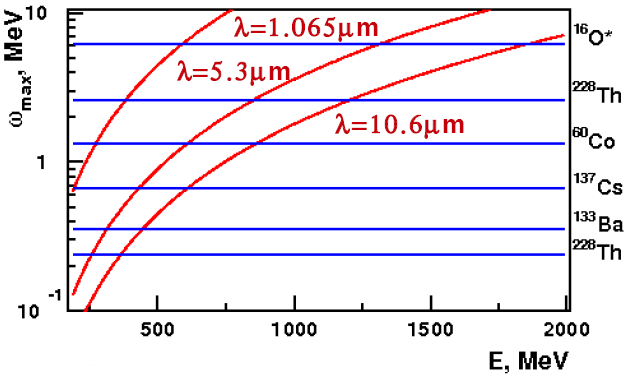}
\caption{Relation between $\omega_{max}$ and  $E$ for different
laser wavelengths. The solid lines are the energies of $\gamma$-active
radionuclide reference lines for the HPGe calibration.}
\label{ucmo}
\end{figure}

Laser is the source of initial photons. The main requirements to it are:
\begin{itemize}
\item 
the single generation line,
\item 
high stability of parameters,
\item 
easy maintenance,
\item 
the choice of the wavelength should provide the maximal energy $\omega_{max}$ 
in the range from 0.2 to 6.0 MeV (figure~\ref{ucmo}), for which HPGe 
detector can be calibrated using $\gamma-$active radionuclides 
(table~\ref{tab1}).
\end{itemize}
At the low energy region $E<0.5$ GeV the solid state  laser with wavelength
$\lambda\approx 1.065$ $\upmu$m can be used. At VEPP-2000 to measure the beam
energy below 1 GeV the PL3 CO laser from Edinburgh Instruments with power of
2~W at $\lambda\approx 5.3$ $\upmu$m is used \cite{ems2000}. The GEM Selected
50$^{TM}$ CO$_2$ lasers from Coherent, Inc with $\lambda\approx 10$ $\upmu$m 
and power of 25 W are implemented to determine the
beam energy $1<E<2$ GeV at VEPP-4M and BEPC-II colliders \cite{vepp4,bems}. 

\subsection{Laser-to-vacuum insertion system}

\begin{figure}[tbp]
\centering
\includegraphics[width=1.0\textwidth]{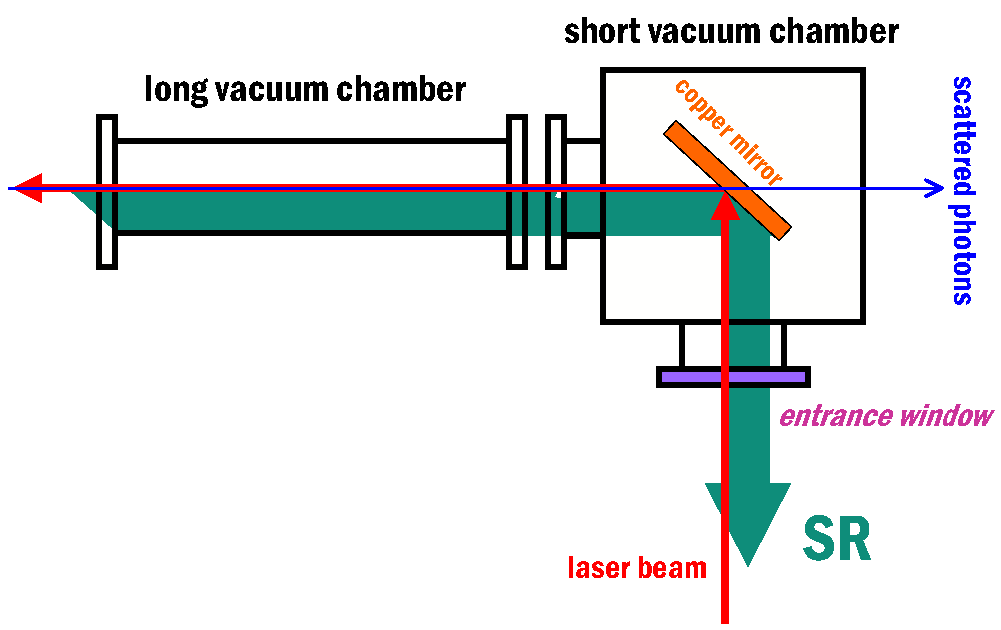}
\caption{Simplified schematic of the laser-to-vacuum insertion assembly.}
\label{cxe-bakyy}
\end{figure}

The insertion of the laser beam into the collider vacuum chamber is performed
using the laser-to-vacuum insertion system. The system is the special
stainless steel vacuum chamber with a  entrance viewport and water cooled 
copper mirror (figure~\ref{cxe-bakyy}) \cite{vacsys,bems}. The system provides 
extra-high vacuum  -- pressure of residual gas less then $5\times 10^{-10}$ 
Torr. The viewport transfers laser light and some amount of synchrotron 
radiation (SR) light to
monitor the beam position. In the vacuum chamber, the laser beam is reflected 
through an angle of $90^\circ$ by the copper mirror.  The mirror
can be turned by bending the vacuum flexible bellows, so the angle between the
mirror and the laser beam can be adjusted as necessary. SR photons heat the 
mirror. The extraction of heat is provided by a water cooling system.
After back-scattering, the photons return to the mirror, pass through it, 
leave the  vacuum chamber, and are detected by the HPGe detector. Note, the 
copper mirror protects the viewport against high power SR 
due to low reflectivity of high energy photons (less than 1\%) from a metallic
surface.

There are two types of viewports. The first one is based on GaAs mono-crystal
plate with diameter of 50.8 mm and thickness of 3 mm. The second is based on
ZnSe  polycristal plate with diameter of 50.8 mm and thickness of 8 mm. Both
viewports are manufactured using similar technology \cite{viewport,bems} and
provide:
\begin{enumerate}
\item
baking out the vacuum system up to 250$^\circ$C,
\item
extra-high vacuum,
\item
 transmission spectrum from 0.9 up to 18 $\upmu$m (GaAs viewport) and from 0.45
 to 20 $\upmu$m (ZnSe viewport).
\end{enumerate}
The advantage of ZnSe viewport is that it is transparent for the visible part
of SR light. This makes the BEMS adjusting more convenient.

\subsection{Optical system.}

\begin{figure}[tbp]
\centering
\includegraphics[width=1.0\textwidth]{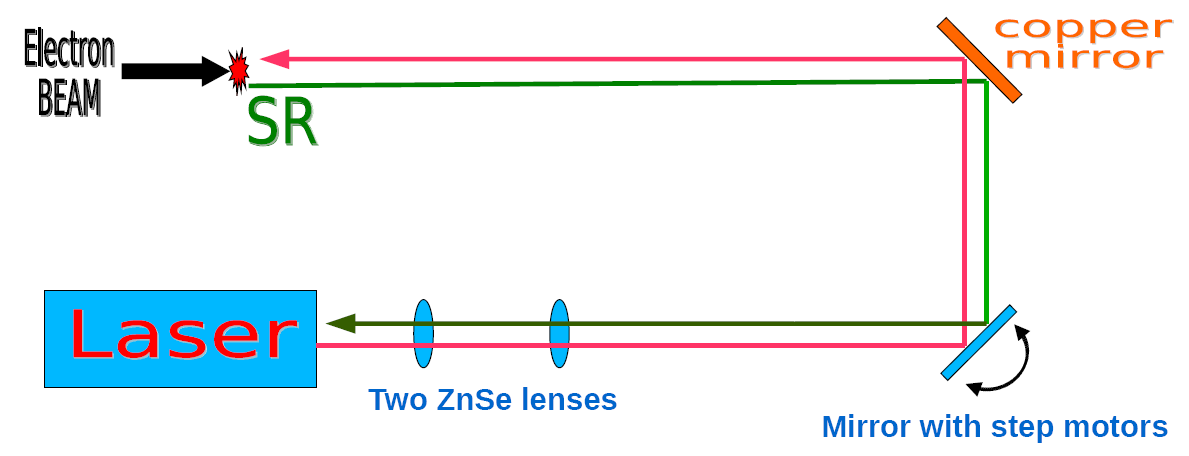}
\caption{Simplified schematic of the optical system.}
\label{optsys}
\end{figure}

The optical system includes the following main units (figure~\ref{optsys}):
\begin{itemize}
\item
Two ZnSe lenses which focus a laser beam at $e-\gamma$ interaction region.
\item
The mirror which reflects the laser beam to a viewport of insertion system. It
is installed on a special support that allows precise vertical and
horizontal angular alignment by using stepping motors.
\end{itemize}
The optical elements of the system are aligned using the SR light of the 
electron beam in such a way that the SR light comes to the laser output window.

\subsection{Data acquisition system.}

\begin{figure}[tbp]
\centering
\includegraphics[width=1.0\textwidth]{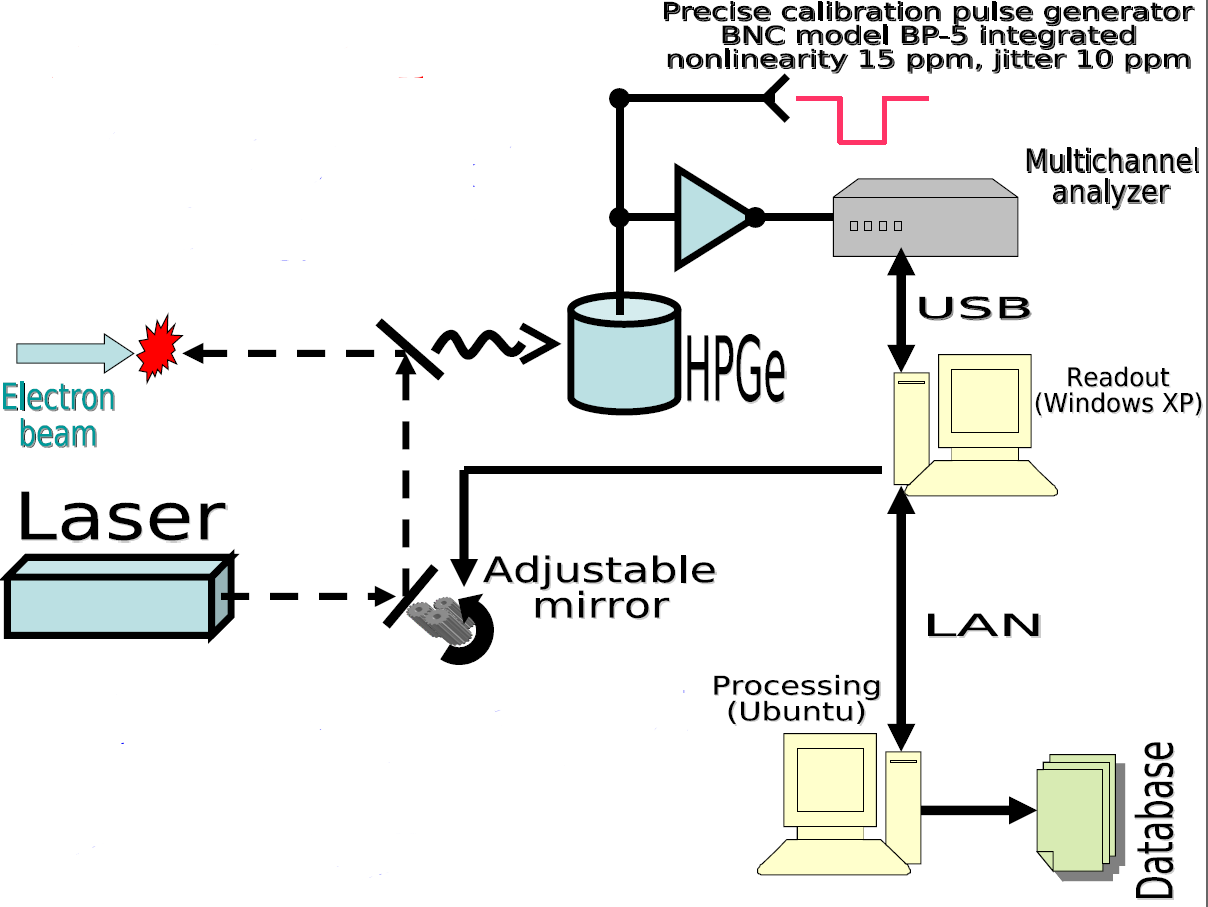}
\caption{Layout of data acquisition system.}
\label{daq}
\end{figure}

The BEMS data acquisition system is shown in figure~\ref{daq}. The MCA digitizes
the signal from the HPGe detector and produces the energy spectrum. It is 
connected to a Windows PC. All spectra processing, monitoring and control 
over the devices involved in the BEMS is concentrated in another PC, under the
control of Linux.

The data acquisition procedure is as follows. The HPGe detector
measurements are read every few seconds, and the detector counting
rate is calculated. If the requested acquisition time has elapsed, or
if conditions of the spectrum acquisition changed sufficiently,
the current spectrum is saved to a file and the next spectrum acquisition 
cycle is launched. Simultaneously, another process periodically requests 
information from collider database and writes the collider parameters, such 
as beam currents, lifetimes, etc to the file.

The data for the HPGe detector calibration -- peaks of the $\gamma$ sources and
peaks of the precise calibration pulser BNC model BP-5 with
integrated nonlinearity $\pm 15$ ppm and jitter $\pm 10$ ppm are accumulated
simultaneously with scattered photons. Generator signals are put to the
preamplifier with 12 different amplitudes covering the range of MCA and
frequency of about 40 Hz. The pulse shape is set in such a way that it is similar to
the shape of the signal from a $\gamma$-quantum.

After finishing the spectrum acquisition cycle, another program
processes the spectrum; it calibrates the energy scale, finds the
Compton edge, and calculates the beam energy. The beam energy is
written into the  database.  

During data taking, mirror are adjusted automatically to provide
maximal photon/electron interaction efficiency, using the
feedback from the detector counting rate.  The processing of the beam energy
measurement is fully automated.

\subsection{Data processing.}

\begin{figure}[tbp]
\centering
\includegraphics[width=1.0\textwidth]{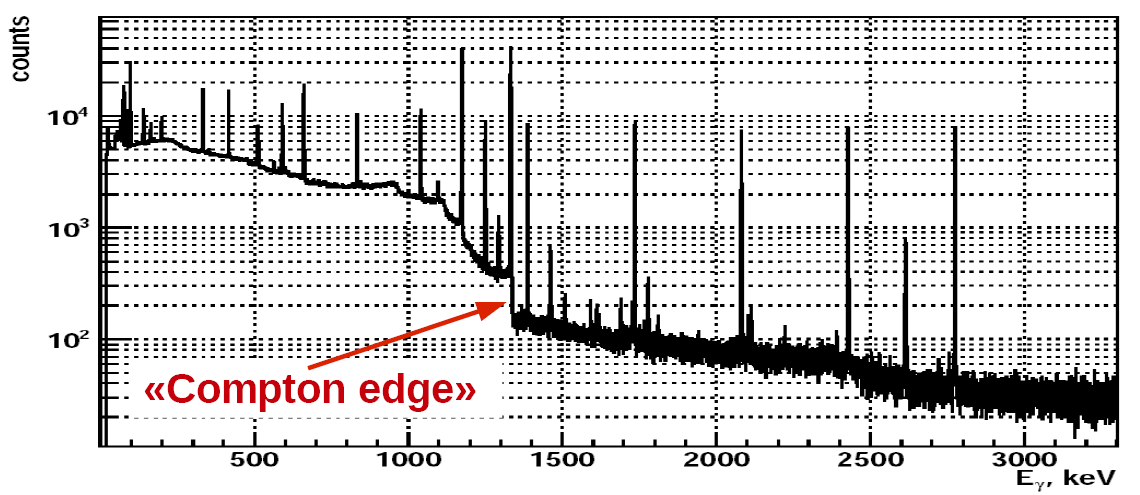}
\caption{The energy spectrum detected by HPGe detector at VEPP-2000. Peaks
correspond to the calibration generator and monochromatic
$\gamma-$radiation sources.}
\label{cnekmp-1}
\end{figure}

The processing of the spectrum (figure~\ref{cnekmp-1}) includes calibration of 
the energy scale, Compton edge fitting and determination of the beam energy.

\begin{figure}[tbp]
\centering
\includegraphics[width=1.0\textwidth]{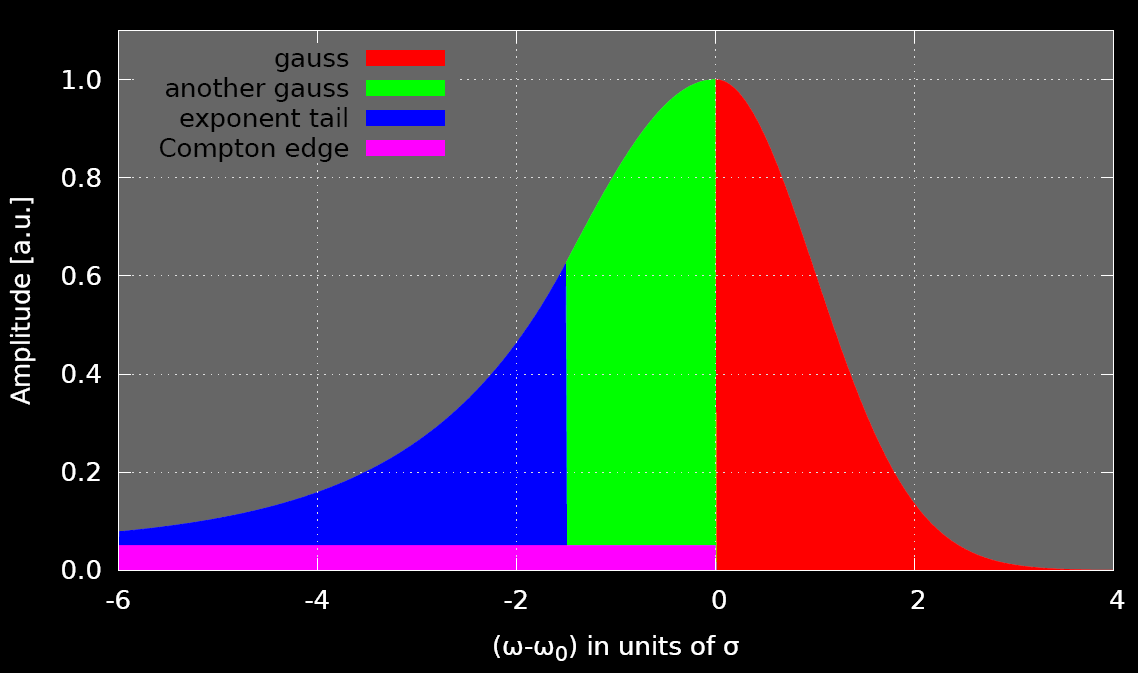}
\caption{HPGe detector response function.}
\label{responce-fun}
\end{figure}

The goal of the HPGe detector calibration is to obtain the coefficients
needed for conversion of the MCA counts  into the
corresponding energy deposition, measured in units of keV, as well as
to determine the parameters of the detector response function. The
following response function is used (figure~\ref{responce-fun}):
\begin{equation}
f(x,x_0) = M \cdot
\left\{\begin{array}{ll}
\exp\biggl\{ {-{(x-x_0)^2\over 2\sigma^2}} \biggr\},
& 0< x+x_0<+\infty, \\
C+(1-C)\exp\biggl\{ {{(x-x_0)^2}\over{ 2(K_0\sigma)^2}} \biggr\},
& -K_0K_1\sigma < x-x_0\leq 0,  \\
C+(1-C)\exp\biggl\{ K_1\biggl({{x-x_0}\over{ K_0\sigma}}+{K_1\over 2}
\biggr) \biggr\},
& -\infty < x-x_0\leq -K_0K_1\sigma,  \\
\end{array}\right.
\label{m-gauss-norm}
\end{equation}
where $M$ is normalization, $x_0$ is the position of the maximum, $\sigma$ and
$K_0\sigma$ are RMS of the Gaussian distribution to the right and to the left
of the $x_0$, respectively, $C$ is responsible for the small-angle Compton
scattering of
$\gamma$-quanta in the passive material between the source and the detector,
$K_1$ is an asymmetry parameter.

\begin{figure}[tbp]
\centering
\includegraphics[width=1.0\textwidth]{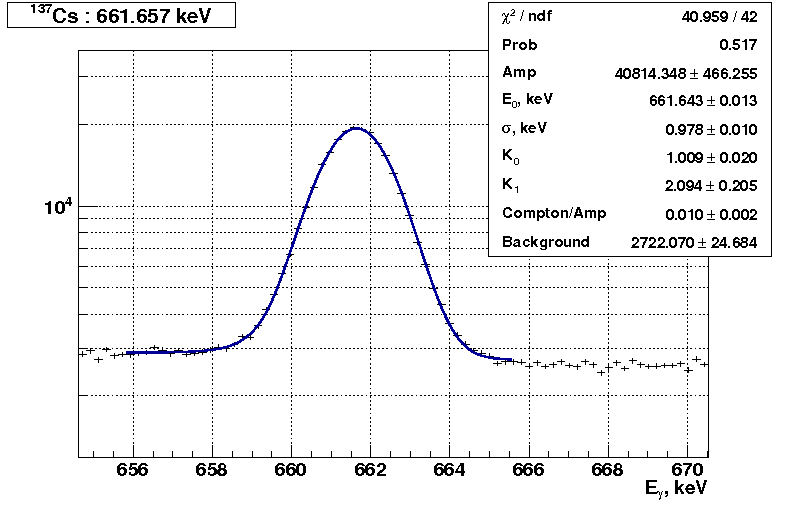}
\caption{The fit to the $^{137}Cs$ 661 keV peak.}
\label{cnekmp-3}
\end{figure}

The calibration procedure is as follows \cite{bems,ems2000}:
\begin{enumerate}
\item
Peak search and identification of the calibration lines.
\item
The calibration peaks are fitted by the sum of response function and
background (figure~\ref{cnekmp-3}).
\item
Using generator data the nonlinearity of MCA scale is obtained.
\item 
Using the results of the isotope peak approximation, the energy dependence of
the response function parameters $\sigma$, $K_0$, $K_1$ and $C$ are determined.
\end{enumerate}

The width of backscattered photons spectrum edge depends on the HPGe 
detector resolution and the electron beam energy spread. 
The spectrum edge is fitted by the function, which takes into account the
<<pure>> edge shape, detector response function, energy spread of scattered
photons due to the energy distribution of the collider beam. The edge
position $\omega_{max}$ and the photons energy spread $\sigma_\omega$ are
obtained from the fit. 
The beam energy $E$ and energy 
spread $\sigma_E$ are calculated from $\omega_{max}$ and $\sigma_\omega$.

\section{Performance of BEMS at collider experiments.}

\subsection{BEMS at VEPP-4M.}

\begin{figure}[tbp]
\centering
\includegraphics[width=1.0\textwidth]{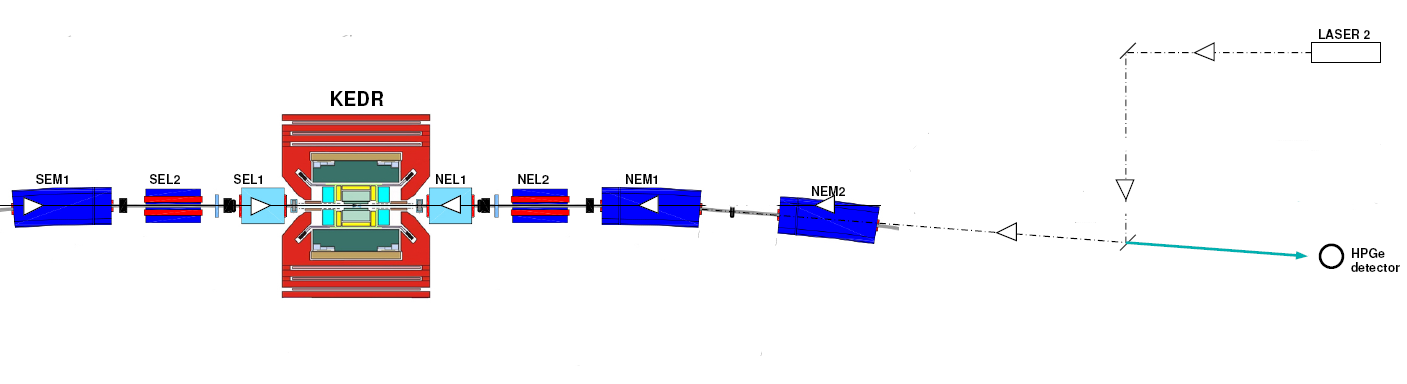}
\caption{Layout of VEPP-4M beam energy measurement system.}
\label{bems-vepp4}
\end{figure}

The accurate beam energy determination is essential for precise mass 
measurements of $J/\psi,$ $\psi^\prime, $ $\psi(3770),$ $D$ mesons and 
$\tau-$lepton in
experiments with the KEDR detector \cite{kedr} at the electron-positron 
collider VEPP-4M \cite{vepp4m}. The RD technique provides precise
instantaneous energy calibration. CBS method allows continuous on-line
monitoring of the beam energy. The layout of BEMS is shown at
Figure~\ref{bems-vepp4}. The laser and electron beams interact in the straight
section of the collider's ring near KEDR detector. Since the electron and
positron beams circulate in the same vacuum chamber in the same magnetic
field their energies are identical. The source of initial
photons is CO$_2$ laser and $\omega_{max}=2 - 7$ MeV for $E=1$ -- 2 GeV.

\begin{figure}[tbp]
\centering
\includegraphics[width=1.0\textwidth]{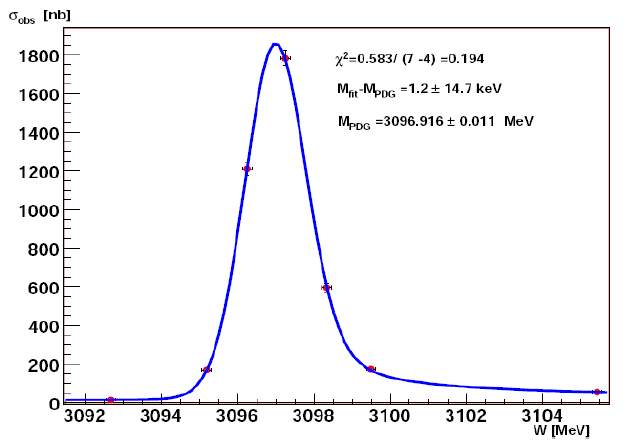}
\caption{Fit to the $J/\psi$ at VEPP-4M.}
\label{jpsikedr}
\end{figure}

The test of CBS method accuracy was performed by comparison of precisely
known $J/\psi$ resonance mass $3096.916\pm0.011$ \cite{pdg} MeV with its value
obtained using BEMS (figure~\ref{jpsikedr}). The mass difference was found to be 
$\Delta m=1.2\pm 14.7$ keV. Deviation of the measured beam energy
from the actual value can be estimated as $\Delta E=\Delta m/2=0.6\pm 17.4$
keV, then relative accuracy of the beam energy determination is
$\delta E/E=5\times 10^{-6}$.

\begin{figure}[tbp]
\centering
\includegraphics[width=1.0\textwidth]{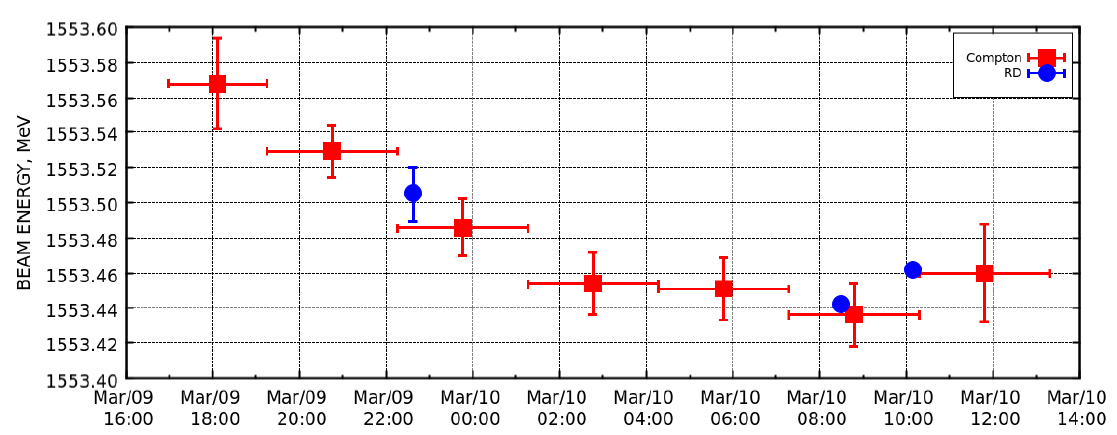}
\caption{Comparison of RD (dots) and CBS (squares) measurements at
the beam energy $E=1553.4$ MeV.}
\label{rd-cbs-vepp4-a}
\end{figure}

Direct comparison of CBS and RD methods was performed at the beam energy
$E=1553.4$ MeV and $E=1884$ MeV. At the beam energy 1553.4 MeV there was no
systematical bias between RD and CBS results within statistical errors of the
CBS measurements  (figure~\ref{rd-cbs-vepp4-a}) and 
$\delta E/E=1.3\times 10^{-5}$. At the beam energy 1884 MeV the difference 
between RD and CBS measurements was $13\pm 38$ keV, and
$\delta E/E=2\times 10^{-5}$. Using results of the  tests the relative error 
of the CBS  measurements at VEPP-4M can be estimated as $10^{-5}$.

\subsection{BEMS at BEPC-II.}

\begin{figure}[tbp]
\centering
\includegraphics[width=1.0\textwidth]{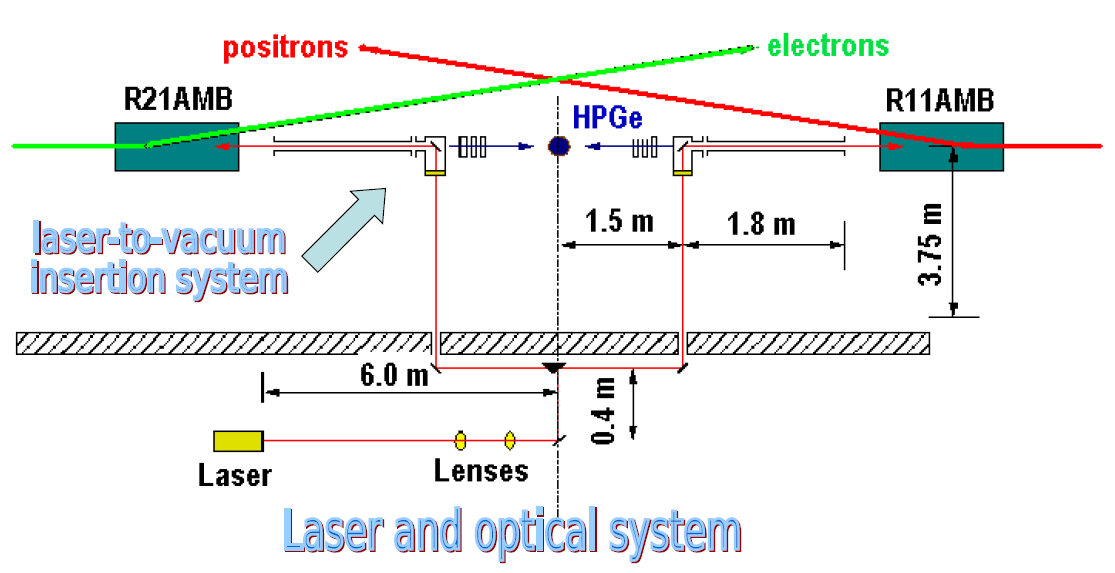}
\caption{Layout of BEPC-II beam energy measurement system.}
\label{layout-bepc}
\end{figure}

The BEMS (figure~\ref{layout-bepc}) is located at the north interaction point of 
BEPC-II collider \cite{bepc}, while the BESIII detector \cite{bes3} is 
installed at the south 
interaction point. This location provides measurement of the electron and 
positron beams energy by the same  HPGe detector. The laser and electron
(positron) beams interact in the straight sections of the collider's rings 
beyond the R2IAMB (R1IAMB) dipole magnets. The source of initial photons is 
CO$_2$ laser. The laser beam is directed to either the electron or  positron 
beams. The energy of the electron and positron beams are measured one after
another, in turn. The beam energy in the south interaction point is
calculated using measured energy, tacking into account the energy losses due
to synchrotron radiation.

Since the HPGe detector is located near the collider's beam pipes,
background due to beam loss is extremely high.  In order to protect
the HPGe detector from background, it is surrounded by 5 cm of lead
on the sides, by 1.5 cm of iron below, and by 5 cm of lead above. The
detector is also shielded by 10 cm of paraffin on all sides. Since
the main background comes from the beam direction, an additional 11
cm of lead is installed in these directions. Another 10 cm of lead
can be inserted into the beam using movable stages to shield from the
beam direction that is not being measured and moved out when the
beam is being measured.
	 
The  systematical accuracy was studied by comparison of the well known mass of
the $J/\psi$ and $\psi^\prime$ resonance with its value obtained using the 
BEMS. One comparison was done after the BEMS was put to operation in
2009 and another during energy scan near the $\tau$ pair threshold in 2010
\cite{bems,tau}. The results are presented in table~\ref{tabscan}. The
relative accuracy of the beam energy determination can be estimated as
$5\times 10^{-5}$.
\begin{table}[tbp]
\caption{The deviations of the beam energy from the actual value at BEPC-II.}
\label{tabscan}
\smallskip
\centering
\begin{tabular}{|lc|c|}
\hline
Scan & $\Delta E$, keV & $\delta E/E$ \\
$J/\psi$(2010)       &$74  \pm 57$ & $6\times 10^{-5}$ \\
$\psi^\prime$(2010)  &$118 \pm 79$ & $7\times 10^{-5}$ \\
$\psi^\prime$(2009)  &$1   \pm 36$ & $2\times 10^{-5}$ \\
\hline
\end{tabular}
\end{table}

\subsection{BEMS at VEPP-2000.}

\begin{figure}[tbp]
\centering
\includegraphics[width=1.0\textwidth]{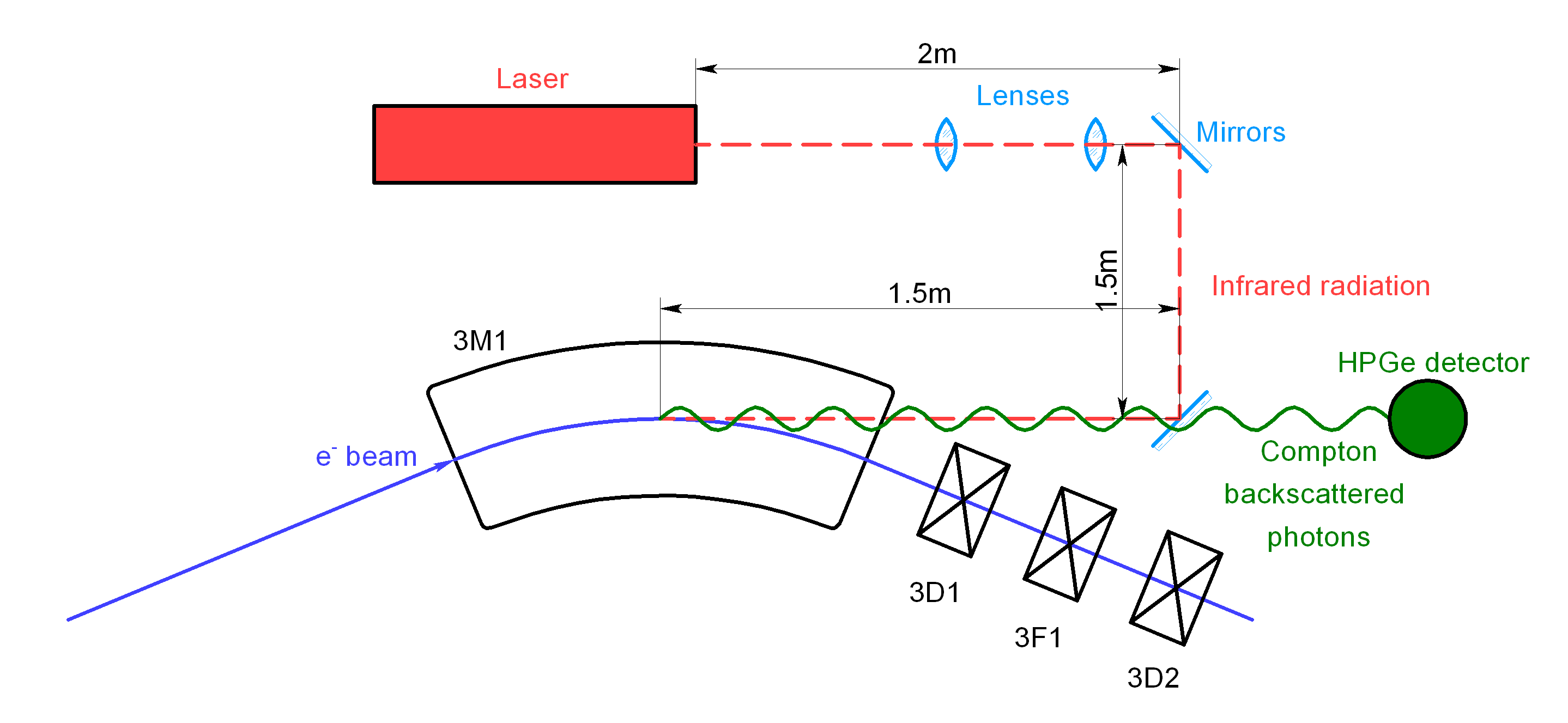}
\caption{Layout of VEPP-2000 beam energy measurement system.}
\label{bems2000}
\end{figure}

\begin{figure}[tbp]
\centering
\includegraphics[width=1.0\textwidth]{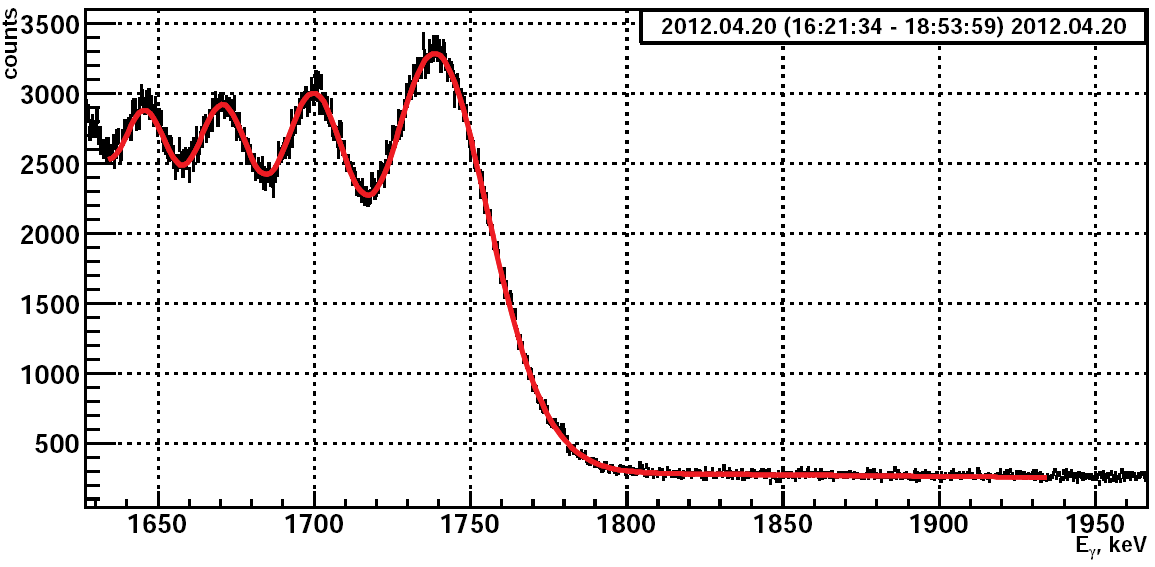}
\caption{The edge of energy spectrum with the fit at VEPP-2000.}
\label{spectr2000}
\end{figure}

At VEPP-2000 the interaction of laser photons with electrons occurs inside
bending magnet ($\rho=140$ cm) at the curvilinear part of the orbit 
(figure~\ref{bems2000}). The source of initial photons is CO laser such that
$\omega_{max}= 0.2 - 2$ MeV for the beam energy $E<1$ GeV. In this case the 
spectrum of scattered photons (figure~\ref{spectr2000}) differs from that 
expected from
by the Klein-Nishina cross  section and scattering kinematics of free 
electrons \cite{prl}. The interference of scattered photons is observed in the
energy spectrum. The systematic error of the beam energy determination was 
tested by comparison with a measurement using the resonance depolarization 
method at the beam energy 458 and 509 MeV and is estimated as 
$6\times 10^{-5}$.

\section{CBS method for the beam energy above 2 GeV.}

Application of the CBS method is constrained by the requirement
$\omega_{max}\lesssim 10$ MeV, i.e. by the laser wavelength. The BEMS for the beam 
energy above 2 GeV can be based on the far infrared laser (FIR) with 
wavelength $\lambda\sim 100$ $\upmu$m. Such a device FIRL 100 is manufactured 
by Edinburgh Instruments with maximal power of 150 mW at $\lambda\approx 119$
and $184$ $\upmu$m. For these wavelength the $\omega_{max}\approx 0.2 - 10$ MeV
at the beam energy $E=2 - 8$ GeV.
At SPring-8 synchrotron radiation facility the MeV photons was produced by
the CBS of the FIR laser radiation with $\lambda=119\mu$m and power of 1.6 W
at the 8 GeV electron beam \cite{spring8}.

\section{Conclusion.}

The CBS method is effective tool for collider energy measurement and
monitoring. The method can be applied for the electron beam energy below 2
GeV. The relative accuracy of the method is $10^{-4} - 10^{-5}$. For
determination of the beam energy from 2 to 8 GeV the FIR laser can be used,
but this requires additional studies.

\acknowledgments

The work was supported by the
Ministry of Education and Science of the Russian Federation, by the RF
Presidential Grant for Scientific Schools NSh-2479.2014.2 and by the RFBR
Grants No 13-02-00418-a, No 14-02-00129-a. The authors are grateful to Z.K.
Solagadze for useful discussions.

\end{document}